\documentclass[aps,twocolumn,floatfix,%showpacs,showkeys,
preprintnumbers,nofootinbib,superscriptaddress]{revtex4}

\usepackage{graphicx}
\usepackage{amsmath}
\usepackage{amsfonts}
\usepackage{amssymb}
\usepackage{color}
\usepackage{subfigure}
\usepackage{epsfig}
\usepackage{morefloats}
\usepackage{multirow}
\usepackage{graphicx,booktabs}
\usepackage{mathrsfs}
\usepackage{txfonts}
\usepackage{indentfirst}
\usepackage{graphicx,booktabs}

\usepackage{longtable,lscape}
\usepackage[colorlinks, citecolor=blue,anchorcolor=red,menucolor=red, linkcolor=red,filecolor=red,urlcolor=blue,frenchlinks=red]{hyperref}

\newcommand{\vsig}{\mbox{\boldmath$\sigma$\unboldmath}}

\usepackage{comment}

\begin{document}

%\begin{spacing}{2.0}

%\title{Toward discovering the excited $\Omega$ baryons through nonleptonic decays processes
%$\Omega_b \rightarrow J/\psi \Omega^*(X)$}
%\title{Toward the discovery of excited $\Omega$ hyperons through charmful $\Omega_b$ weak decays}
\title{Excited $\Omega$ hyperon in charmful $\Omega_b$ weak decays}
%nonleptonic decay processes $\Omega_b \rightarrow J/\psi \Omega(X)$}

\author{Kai-Lei Wang}
\email{wangkaileicz@foxmail.com}
\affiliation{Department of Physics,
Changzhi University, Changzhi, Shanxi 046011, China}
\affiliation{Synergetic Innovation Center for Quantum Effects and Applications (SICQEA),\\
Hunan Normal University, Changsha 410081, China}

\author{Juan Wang}
\email{wjuanmm@163.com}
\affiliation{Department of Physics,
Changzhi University, Changzhi, Shanxi 046011, China}
\affiliation{School of Physics and Information Engineering,
Shanxi Normal University, Taiyuan 030031, China}

\author{Yu-Kuo Hsiao}
\email{yukuohsiao@gmail.com}
\affiliation{School of Physics and Information Engineering,
Shanxi Normal University, Taiyuan 030031, China}

\author{Xian-Hui Zhong}
\email{zhongxh@hunnu.edu.cn}
\affiliation{Synergetic Innovation Center for Quantum Effects and Applications (SICQEA),\\
Hunan Normal University, Changsha 410081, China}
\affiliation{Department of Physics, Hunan Normal University, and Key
Laboratory of Low-Dimensional Quantum Structures and Quantum Control
of Ministry of Education, Changsha 410081, China}

\date{\today}

\begin{abstract}
We investigate the sextet $b$-baryon decay processes $\Omega_b\to J/\psi\Omega^{(*)}$,
where $\Omega^*$ represents the $1P$-, $1D$- and $2S$-wave excited $\Omega$ hyperons
in the spectroscopy. Using the constituent quark model, we obtain
${\cal B}(\Omega_b \to J/\psi\Omega)=8.8\times 10^{-4}$, which agrees with the previous studies
to the order of magnitude. By identifying $\Omega(2012)$ as $\Omega(1^2P_{3/2^-})$,
${\cal B}(\Omega_b \to J/\psi\Omega(2012))=1.1\times 10^{-3}$ can be similarly significant.
Additionally, $\Omega(1^2D_{5/2^+})$ and $\Omega(1^4D_{3/2^+},1^4D_{5/2^+})$ states
exhibit production rates of 0.5, and (0.6, 0.8), respectively,
relative to their ground-state counterpart. Notably, our findings suggest that
${\cal B}(\Omega_b\to J/\psi\Omega(2^2S_{1/2^+},2^4S_{3/2^+}))$ are as large as
$(4.5,20)\times 10^{-4}$, making them accessible to experiments at LHCb.
\end{abstract}

\maketitle
\section{Introduction}
In baryon spectroscopy, the formation of a colorless state from three quarks
with three different colors is fundamentally linked to
the principles of quantum chromodynamics (QCD) and hadron physics.
While the combination of individual quark spins and orbital angular momenta
is expected to produce a variety of
light baryon states~\cite{Klempt:2009pi,Crede:2013kia,Forkel:2008un},
only a subset of these states has been observed to date~\cite{pdg}.
The so-called {\it ``missing excited baryon''} problem
highlights an incomplete theoretical understanding
of baryon spectroscopy~\cite{Capstick:2000dk,Klempt:2009pi}.

The $\Omega$ hyperon spectroscopy has been relatively underexplored. Nonetheless,
Belle initiated a new era of discoveries with the identification of an excited hyperon,
$\Omega(2012)$~\cite{Belle:2018mqs}, which was later reconfirmed through the process
$\Omega_c^0 \to \pi^+ \Omega(2012)^-,\Omega(2012)^-\to \Xi^0K^-$~\cite{Belle:2021gtf}.
Additionally, a heavier excited hyperon $\Omega(2109)^-$, observed via
$e^+ e^-\to \Omega(2109)^-\bar \Omega^+ +c.c.$, and reported by BESIII~\cite{BESIII:2024eqk}.
To investigate the nature of these new excited hyperons,
significant theoretical attention has been devoted~\cite{Xiao:2018pwe,Liu:2019wdr,Wang:2018hmi,
Zeng:2020och,Aliev:2018yjo,Aliev:2018syi,Arifi:2022ntc,Polyakov:2018mow,Hu:2022pae,
Ikeno:2022jpe,Lin:2019tex,Lu:2020ste,Ikeno:2020vqv,Xie:2021dwe,
Valderrama:2018bmv,Pavao:2018xub,Huang:2018wth,Gutsche:2019eoh}.
Notably, $\Omega(2012)$ is often preferred
as an exotic molecule candidate~\cite{Hu:2022pae,Ikeno:2022jpe,Lin:2019tex,Lu:2020ste,
Ikeno:2020vqv,Xie:2021dwe,Valderrama:2018bmv,Pavao:2018xub,Huang:2018wth,Gutsche:2019eoh}
rather than as a conventional $1P$-wave $\Omega(sss)$ state.

Clearly, the {\it ``missing excited baryon''} problem persists in $\Omega$ hyperon spectroscopy,
as all waves of $\Omega$ hyperons have been theoretically
predicted~\cite{Oh:2007cr,Capstick:1986ter,Faustov:2015eba,
Loring:2001ky,Liu:2007yi,Chao:1980em,Chen:2009de, An:2013zoa,Kalman:1982ut,
Pervin:2007wa,An:2014lga,Engel:2013ig,CLQCD:2015bgi,Carlson:2000zr,Goity:2003ab, Schat:2001xr,Matagne:2006zf,Bijker:2000gq,Aliev:2016jnp},
but only a few have been experimentally observed. These include
$\Omega \equiv\Omega(1672)$~\cite{Abrams:1964tu,Barnes:1964pd},
$\Omega(2012,2109)$~\cite{Belle:2018mqs,Belle:2021gtf,BESIII:2024eqk}, and
$\Omega(2250,2380,2470)$~\cite{Biagi:1985rn,Aston:1987bb,Aston:1988yn}.
Since further clarification and exploration are essential, we propose that
recent measurements of the sextet $b$-baryon decay process
$\Omega_b \to J/\psi\Omega$~\cite{D0:2008sbw,CDF:2009sbo,
LHCb:2013wmn,CDF:2014mon,LHCb:2023qxn,Nicolini:2023stq}
be extended to include $\Omega_b\to J/\psi\Omega^{*}$,
where $\Omega^{*}$ denotes higher-wave excited $\Omega$ hyperons.

\newpage
As LHCb continues to improve the statistics and precision~\cite{LHCb:2023qxn,Nicolini:2023stq},
the absolute branching fractions of $\Omega_b^-$ decays may soon be measured with a similar accuracy
to those of the anti-triplet $b$-baryon decays~\cite{UA1:1991vse,CDF:1992lrw,CDF:2006eul,CDF:1996rvy,
D0:2004quf,D0:2007giz,D0:2011pqa,D0:2012hfl,LHCb:2013hzx,LHCb:2020iux,
LHCb:2019fim,LHCb:2019aci,ATLAS:2014swk,ATLAS:2015hik,CMS:2018wjk,D0:2007gjs}.
Theoretical estimations are therefore needed, and
a variety of theoretical tools are already available for such studies. These include
the non-relativistic quark model~\cite{Fayyazuddin:1998ap,Mott:2011cx,Cheng:1995fe,Cheng:1996cs},
the covariant confined quark model~\cite{Gutsche:2018utw,Gutsche:2017wag,Gutsche:2013oea,Gutsche:2015lea},
the covariant oscillator quark model~\cite{Mohanta:1998iu},
the relativistic three-quark model~\cite{Ivanov:1997hi,Ivanov:1997ra},
the light-front quark model~\cite{Hsiao:2021mlp,Wei:2009np,Zhu:2018jet,Wang:2024mjw},
the perturbative QCD approach~\cite{Rui:2023fiz,Chou:2001bn},
the generalized factorization approach~\cite{Hsiao:2015cda,Fayyazuddin:2017sxq,Hsiao:2015txa},
and SU(3) flavor analysis~\cite{Dery:2020lbc}.

In this work, we employ the constituent quark model~\cite{Niu:2020gjw,Niu:2025lgt,Wang:2022zja,
Niu:2021qcc,Niu:2020aoz,Pervin:2006ie,Pervin:2005ve} to provide our estimations.
As we will demonstrate, this model can be extended to study both
the $\Omega_b\to J/\psi\Omega$ and $\Omega_b\to J/\psi\Omega^{*}$ decay processes,
enabling a systematic analysis of the missing excited $\Omega$ states.
This paper is organized as follows:
In Sec.~\ref{MODEL}, we apply the constituent quark model to two-body nonleptonic
weak decays of $\Omega_b \to J/\psi\Omega^{(*)}$.
In Sec.~\ref{numerical}, we present the numerical results.
Finally, we provide our discussions and conclusion in Sec.~\ref{discussions}.

\section{framework}\label{MODEL}
\begin{figure}[t]
\centering
\includegraphics[width=0.48\textwidth]{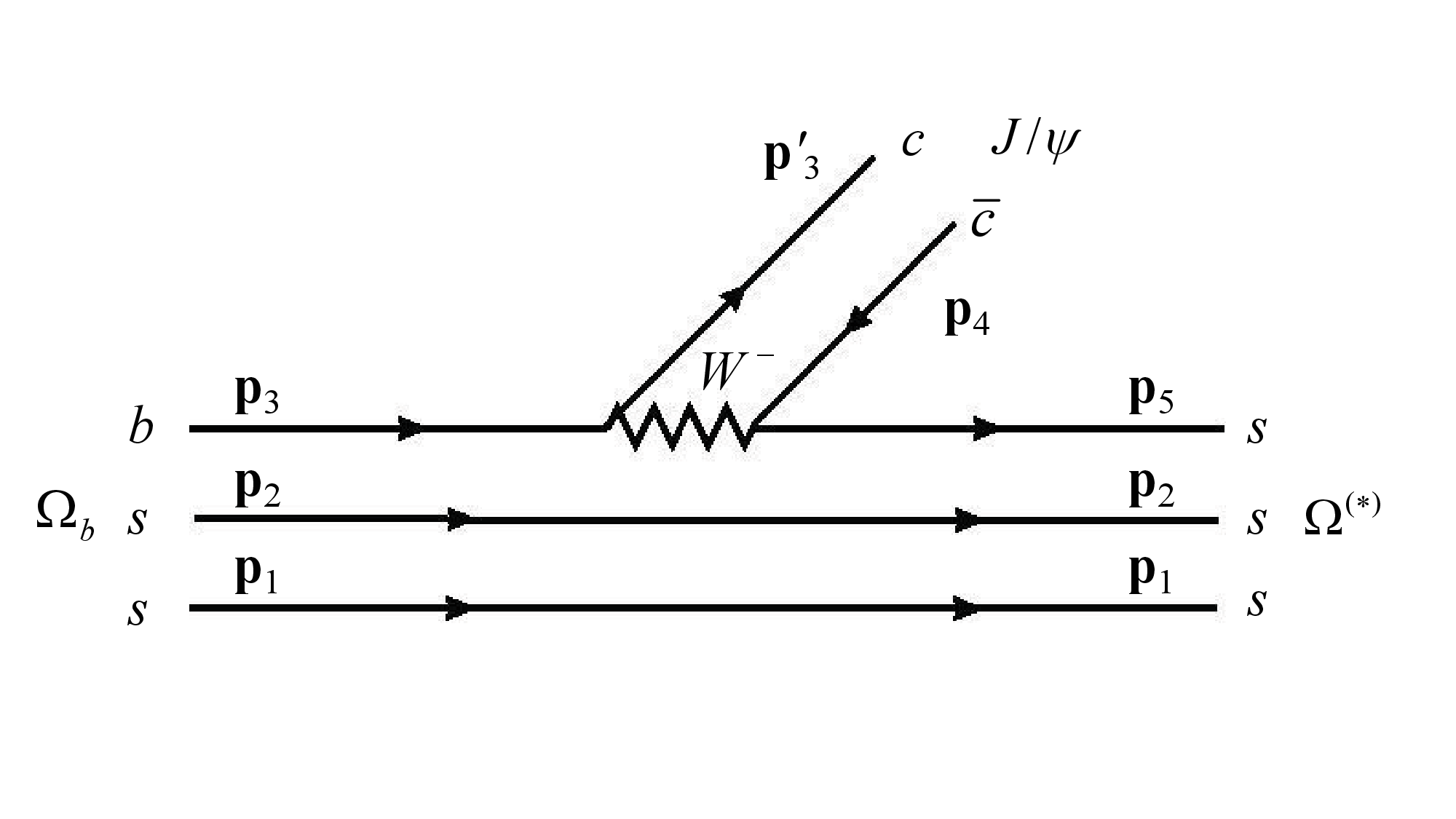}    \vspace{-1.5 cm}
\caption{Feynman diagram for the nonleptonic weak decay
$\Omega_b\rightarrow J/\psi\Omega^{(*)}$.}\label{tu}
\end{figure}

As shown in Fig.~\ref{tu}, the recently observed sextet $b$-baryon decay, $\Omega_b \to J/\psi\Omega$,
proceeds exclusively via the internal $W$-boson emission diagram~\cite{Cheng:2021qpd},
where $\Omega$ is strait-forwardly formed through the $\Omega_b\to\Omega$ transition.
The single decay topology, extended to $\Omega_b\to J/\psi\Omega^{*}$,
also facilitates the investigation of the $1P$-, $2S$- and $1D$-wave excited $\Omega^{*}$ states.

The quark-level effective Hamiltonian of $b\to c\bar{c}s$ weak transition
induces the doubly charmful decay channels
$\Omega_b \to J/\psi\Omega^{(*)}$,
given by~\cite{Buchalla:1995vs}
\begin{eqnarray}\label{dww}
H_W&=&\frac{G_F}{\sqrt{2}}V_{bc} V_{cs}^* (C_1\mathcal{O}_1+C_2\mathcal{O}_2)\,,
\end{eqnarray}
where $G_F$ is the Fermi constant,
$C_1$ and $C_2$ are the Wilson coefficients, and ($V_{bc}$, $V_{cs}$) are
the Cabibbo-Kobayashi-Maskawa (CKM) matrix elements.
Additionally, ${\cal O}_{1,2}$ are the current-current operators,
which are written as
\begin{eqnarray}\label{dww}
\mathcal{O}_1&=&\bar{\psi}_{\bar{s}_\beta}\gamma_\mu(1-\gamma_5)\psi_{c_\beta}\bar{\psi}_{\bar{c}_\alpha}\gamma^\mu(1-\gamma_5)\psi_{b_\alpha}\,,\nonumber\\
\mathcal{O}_2&=&\bar{\psi}_{\bar{s}_\alpha}\gamma_\mu(1-\gamma_5)\psi_{c_\beta}\bar{\psi}_{\bar{c}_\beta}\gamma^\mu(1-\gamma_5)\psi_{b_\alpha}\,.
\end{eqnarray}
Here, $\psi_{j_\delta}$ is the $j$th quark field, where $j$ can be $s$, $c$ or $b$,
and $\delta=(\alpha,\beta)$ is the color index.
We  define the convention for the quark and antiquark fields
\begin{equation}
\small{
\begin{aligned}
\psi_c(x) = \int \frac{d p}{(2\pi)^{3/2}} \left(\frac{m}{p^0}\right)^{1/2} \sum_{s} \left[u_s(p) b_{s,c}(p) e^{-i p \cdot x} + v_s(p) d^\dagger_{s,c}(p) e^{i p \cdot x}\right],\\
\bar{\psi}_c(x) = \int \frac{d p}{(2\pi)^{3/2}} \left(\frac{m}{p^0}\right)^{1/2} \sum_{s} \left[\bar{u}_{s}(p) b^\dagger_{s,c}(p) e^{i p \cdot x} + \bar{v}_s(p) d_{s,c}(p) e^{-i p \cdot x}\right].
\end{aligned}}
\end{equation}
where $c$  is color quantum number, and
\begin{eqnarray}\label{Spinorfield}
u_s(p)=\sqrt{\frac{E+m}{2m}} \left(\begin{array}{c} \varphi_{s} \cr \frac{\mathbf{\sigma} \cdot \emph{\textbf{p}}}{E+m}\varphi_{s} \end{array}\right),
v_s(p)=\sqrt{\frac{E+m}{2m}} \left(\begin{array}{c} \frac{\mathbf{\sigma} \cdot \emph{\textbf{p}}}{E+m}\chi_{s} \cr \chi_{s} \end{array}\right).
\end{eqnarray}
Here, $s$ represents spin labeling, so
\begin{eqnarray}\label{Spinorfield}
\sum_s \varphi_s\varphi_s^{\dag}=\sum_s \chi_s\chi_s^{\dag}=1.
\end{eqnarray}
The anticommutation relations of the creation and annihilation operators are given by
\begin{eqnarray}
\{b_{s,c}(p),b_{s',c'}^\dagger(p') \}=\{d_{s,c}(p),d_{s',c'}^\dagger(p') \}=\delta_{ss'}\delta_{cc'}\delta^3(p-p').
\end{eqnarray}
The normalization of spinor is
\begin{eqnarray}
\bar{u}_s(p)u_{s'}(p)  = -\bar{v}_s(p)v_{s'}(p) = \delta_{ss'}.
\end{eqnarray}

%=======
%
%=======
%
According to Refs.~\cite{Mannel:1992ti,Buras:1985xv,Blok:1992na}, $C_2 \langle J/\psi \Omega|{\cal O}_2|\Omega_b\rangle$
corresponds to the internal $W$-boson emission diagram. On the other hand,
$C_1\langle J/\psi \Omega |{\cal O}_1|\Omega_b\rangle$ is replaced by
$C_1/N_c^{eff} \langle J/\psi \Omega |{\cal O}_2|\Omega_b\rangle$ in the generalized factorization,
where $N_c^{eff}$ is an effective color number that accounts for the non-factorizable QCD corrections.
In the large $N_c^{eff}$ limit ($N_c^{eff}\to \infty$)~\cite{Mannel:1992ti,Buras:1985xv,Blok:1992na},
the contribution from $C_1\mathcal{O}_1$ becomes negligible.
Therefore, we only consider $C_2\mathcal{O}_2$ in this work.

The constituent quark model separates
$H_W$ into the parity-conserving (PC) and  parity-violating (PV)
components~\cite{Niu:2020gjw}:
\begin{eqnarray}\label{Hww}
H_{W}=H_{W}^{PC}+H_{W}^{PV}\,.
\end{eqnarray}
In the non-relativistic approximation, the two components are
given by~\cite{Niu:2020gjw,Niu:2025lgt,Wang:2022zja}
\begin{eqnarray}\label{Hw}
H_{W}^{PC}&\simeq&\frac{G_F}{\sqrt{2}}V_{cs}^* V_{cb}C_2 \frac{\hat{\phi}_c \hat{O}_f}{(2\pi)^3}\delta^3(\textbf{p}_3-\textbf{p}_3'-\textbf{p}_4-\textbf{p}_5) \nonumber \\
&&\Bigg \{\vsig_4 \cdot \left(\frac{\textbf{p}_5}{2m_5}+\frac{\textbf{p}_4}{2m_4}\right)+ \vsig_3 \cdot \left(\frac{\textbf{p}_3'}{2m_3'}+\frac{\textbf{p}_3}{2m_3}\right)\nonumber \\
& &-\left[\left(\frac{\textbf{p}_3'}{2m_3'}+\frac{\textbf{p}_3}{2m_3}\right)
-i \vsig_3 \times \left(\frac{\textbf{p}_3}{2m_3}-\frac{\textbf{p}_3'}{2m_3'}\right)\right] \cdot \vsig_4    \nonumber \\
&&-\vsig_3 \cdot \left[\left(\frac{\textbf{p}_5}{2m_5}+\frac{\textbf{p}_4}{2m_4}\right)  -i \vsig_4
\times \left(\frac{\textbf{p}_4}{2m_4}-\frac{\textbf{p}_5}{2m_5}\right)\right] \Bigg \}\,,\nonumber\\
H_{W}^{PV}&\simeq&\frac{G_F}{\sqrt{2}}V_{cs}^* V_{cb}C_2 \frac{\hat{\phi}_c  \hat{O}^f}{(2\pi)^3}\delta^3(\textbf{p}_3-\textbf{p}_3'-\textbf{p}_4-\textbf{p}_5)\nonumber\\
& &(\vsig_3 \cdot \vsig_4 - 1),
\end{eqnarray}
%
%
%which are consistent with the representation in Ref.~\cite{Niu:2020gjw,Niu:2025lgt,Wang:2022zja}.
%In above Eq.(\ref{Hw}),
where $\textbf{p}_j$ and $m_j$, as assigned in Fig.~\ref{tu},
are the momentum and mass of the $j$th quark, respectively, and
$\textbf{\vsig}_j$ is the spin operator of the $j$th quark.
Additionally,
$\hat{\phi}_c$ is the color operator and $\hat{O}_f$ is the flavor operator.

Explicitly, we present
\begin{eqnarray}\label{OP1}
%\hat{\phi}_c =&& \frac{1}{\sqrt{3}}[\hat{a}_{c}^{\dag}(R) \hat{a}_{\bar{c}}^{\dag}(\bar{R})+\hat{a}_{c}^{\dag}(B) \hat{a}_{\bar{c}}%^{\dag}(\bar{B})+\hat{a}_{c}^{\dag}(G) \hat{a}_{\bar{c}}^{\dag}(\bar{G})] \nonumber\\
%&&\times [\hat{a}_{s}^{\dag}(R) \hat{a}_{b}(R)+\hat{a}_{s}^{\dag}(B) \hat{a}_{b}(B)+\hat{a}_{s}^{\dag}(G) \hat{a}_{b}(G)]\,,
\hat{O}_f &=&\hat{b}_{3^{\prime}}^\dagger(c)\hat{b}_{4}^\dagger(\bar{c})\hat{b}_5^\dagger(s)\hat{b}_3(b)\,,\nonumber\\
\hat{\phi}_c&=& \delta_{c_4c_{3'}}\delta_{c_5c_3}\,,
\end{eqnarray}
where $\hat{b}^{(\dag)}(q)$ annihilates (creates) the quark (q), and the
$\delta_{c_5c_3}(\delta_{c_4c_{3'}})$  represents that the colors of $c_3(c_{3'})$ and $c_5(c_4)$ should be conserved.

%$\hat{a}_{q}^{(\dag)}(\delta)$ annihilates (creates) the color $\delta$ for $q$.
%%
%The pre-factor $1/\sqrt 3$ in $\hat{\phi}_c^{0\to c\bar c}$, not appearing in the quark current of Eq.~(\ref{dww}),
%is specifically added to normalize the three possible vacuum-induced $\delta\bar \delta$ color pairs.
%$\delta_{c_5c_3}$  represents that the colors of $c_3$ and $c_5$ should be conserved.
%It is interesting to note that the pre-factor has been regarded as a requirement
%for the vacuum to color-anticolor creation. Extensively,
%it can be found in quark model calculations on the vacuum-to-meson form factors~\cite{Chernyak:1981zz,Chernyak:1983ej,Cheng:2009yz,Cheng:2005vd},
%the strong decays of $\Omega_c\to \Xi_c K$~\cite{Garcia-Tecocoatzi:2022zrf,Luo:2023sne,Chen:2017gnu}, and
%the charmonian decays of $M_{c\bar c}\to D\bar D$~\cite{Gui:2018rvv,Deng:2023mza,Yang:2009fj,Wang:2014voa}.

For the color operator to act on
the color wave functions of $\Omega_{b}$, $\Omega^{(*)}$, and $J/\psi$,
we define
\begin{eqnarray}\label{OP2}
&&
|\zeta_{\Omega_b} \rangle
=\frac{1}{\sqrt{6}}(RGB-RBG+GBR-GRB+BRG-BGR)\,,\nonumber\\
&&
|\zeta_{\Omega^{(*)}}\rangle=|\zeta_{\Omega_b} \rangle\,,\nonumber\\
&&
|\zeta_{J/\psi}\rangle=\frac{1}{\sqrt{3}}(R\bar R+G\bar G+B\bar B)\,.
\end{eqnarray}
From Eqs.~(\ref{OP1}) and~(\ref{OP2}),
we obtain
\begin{eqnarray}
\langle \zeta_{J/\psi}\zeta_{\Omega^{(*)}}|\hat{\phi}_c|\zeta_{\Omega_b} \rangle=\sqrt{3}\,.
\end{eqnarray}

In the constituent quark model,
the decay amplitude is represented as
\begin{eqnarray}\label{Hwwb}
\mathcal{M}(\Omega_b\to J/\psi \Omega^{(*)})=
\mathcal{M}^{PC}+\mathcal{M}^{PV}\,,
\end{eqnarray}
where
\begin{eqnarray}\label{Hwwc}
\mathcal{M}^{PC(PV)}& = & \frac{1}{\sqrt{3}}
\langle  J/\psi(\textbf{q};1,J_{J/\psi}^z)\Omega^{(*)}(\textbf{P}_f;J_f,J_f^z)|\nonumber\\
&&(1+\hat{P}_{25}+\hat{P}_{15})H_{W}^{PC(PV)}|\Omega_b(\textbf{P}_i;J_i,J_i^z)\rangle\, .
\end{eqnarray}
Here, the factor $\frac{1}{\sqrt{3}}$ is the normalization coefficient
and $\hat{P}_{ij}$ represents the permutation operator of quarks $i$ and $j$.
In the above equation, $J/\psi(\textbf{q};1,J_{J/\psi}^z)$,
$\Omega_b(\textbf{P}_i;J_i,J_i^z)$, and $\Omega^{(*)}(\textbf{P}_f;J_f,J_f^z)$
are wave functions, where ${\bf q}$ and ${\bf P}_{i(f)}$ are the total momentum,
$J_{i(f)}$ are the total angular momentum,
and $J^z_{i(f)}$ are the third component of the total angular momentum.

It is worth noting that the calculation of the amplitudes for $\Omega_b \to \Omega^{(*)} J/\psi$ decays should consider a question of symmetry.
The flavor wave functions of $\Omega_b$ and $\Omega$ are $\varphi_{\Omega_b} = ssb$ and
$\varphi_{\Omega} = sss$, respectively. In the $\Omega_b \rightarrow \Omega^{(*)} J/\psi$  nonleptonic decay process,  the bottom quark of
the $\Omega_b$ decays into an $s$ quark, which can be any of the three
$s$ quarks in the $\Omega$. Since   the wave functions of the final state $\Omega^{(*)}$ baryon
are fully antisymmetric under the exchange of any two quarks, so we obtain
\begin{eqnarray}\label{Hww}
&  \langle J/\psi(\textbf{q};1,J_{J/\psi}^z)\Omega^{(*)}(\textbf{P}_f;J_f,J_f^z) | H_{W}^{PC(PV)} | \Omega_b(\textbf{P}_i;J_i,J_i^z) \rangle  \nonumber\\
=&\langle J/\psi(\textbf{q};1,J_{J/\psi}^z)\Omega^{(*)}(\textbf{P}_f;J_f,J_f^z)  | \hat{P}_{25}H_{W}^{PC(PV)} | \Omega_b(\textbf{P}_i;J_i,J_i^z) \rangle \nonumber\\
=&\langle J/\psi(\textbf{q};1,J_{J/\psi}^z)\Omega^{(*)}(\textbf{P}_f;J_f,J_f^z)  | \hat{P}_{15}H_{W}^{PC(PV)} | \Omega_b(\textbf{P}_i;J_i,J_i^z) \rangle.\nonumber\\
\end{eqnarray}
Therefore, one can obtain
\begin{eqnarray}\label{Hwwd}
\mathcal{M}^{PC(PV)}& = &\sqrt{3}\langle J/\psi(\textbf{q};1,J_{J/\psi}^z)\Omega^{(*)}(\textbf{P}_f;J_f,J_f^z) | \nonumber\\
&   & H_{W}^{PC(PV)} | \Omega_b(\textbf{P}_i;J_i,J_i^z) \rangle.
\end{eqnarray}
Similar calculations  have been applied in Ref.~\cite{Pervin:2006ie}.

\begin{table}[htbp]
\begin{center}
\caption{\label{wavefunction} The  spin-flavor-space
wave-functions of $\Omega$ baryons under $SU(6)$ quark model classification
are listed below. We denote the baryon states as $|N_{6},^{2S+1}N_{3},N,L,J^{P}\rangle$
where $N_{6}$ stands for the irreducible representation of spin-flavor
$SU(6)$ group, $N_{3}$ stands for the irreducible representation of
flavor $SU(3)$ group and $N$, $L$, $J^{P}$ as principal quantum number,
total orbital angular momentum and spin-parity, respectively\cite{Xiao:2018pwe,Liu:2019wdr,Wang:2018hmi}.
The $\phi$, $\chi$, $\psi$ denote flavor, spin and spatial wave function, respectively.
The Clebsch-Gorden coefficients of spin-orbital coupling and color wave function $\zeta_\Omega$  have been omitted.}
\begin{tabular}{lccccccccccccccccccccccccccccccccccccccccccccc}\hline\hline
%State                            ~~~~ &             ~~~~ &Predicted                ~~~~&Predicted  ~~~~&Predicted   ~~~~&Predicted  ~~~~&Predicted  ~~~~&Predicted ~~~~&        \\
$n^{2S+1}L_{J^p}$  ~~&$|N_6,^{2S+1}N_3,N,L,J^P\rangle$  ~~&Wave function  \\ \hline
$1^4S_{\frac{3}{2}^+}$   ~~&$|56,^{4}10,0,0,\frac{3}{2}^+\rangle$            & $\phi^{s}\chi^{s}\psi_{000}^{s}$                                                               \\
$1^2P_{\frac{1}{2}^-}$   ~~&$|70,^{2}10,1,1,\frac{1}{2}^-\rangle$            & $\frac{1}{\sqrt{2}}\left(\phi^{s}\chi^{\rho}\psi_{11L_{z}}^{\rho}+\phi^{s}\chi^{\lambda}\psi_{11L_{z}}^{\lambda}\right)$   \\
$1^2P_{\frac{3}{2}^-}$   ~~&$|70,^{2}10,1,1,\frac{3}{2}^-\rangle$            & $\frac{1}{\sqrt{2}}\left(\phi^{s}\chi^{\rho}\psi_{11L_{z}}^{\rho}+\phi^{s}\chi^{\lambda}\psi_{11L_{z}}^{\lambda}\right)$  \\
$2^2S_{\frac{1}{2}^+}$   ~~&$|70,^{2}10,2,0,\frac{1}{2}^+\rangle$            & $\frac{1}{\sqrt{2}}\left(\phi^{s}\chi^{\rho}\psi_{200}^{\rho}+\phi^{s}\chi^{\lambda}\psi_{200}^{\lambda}\right)$   \\
$2^4S_{\frac{3}{2}^+}$   ~~&$|56,^{4}10,2,0,\frac{3}{2}^+\rangle$            & $\phi^{s}\chi^{s}\psi_{200}^{s}$                                                                                   \\
$1^2D_{\frac{3}{2}^+}$   ~~&$|70,^{2}10,2,2,\frac{3}{2}^+\rangle$            & $\frac{1}{\sqrt{2}}\left(\phi^{s}\chi^{\rho}\psi_{22L_{z}}^{\rho}+\phi^{s}\chi^{\lambda}\psi_{22L_{z}}^{\lambda}\right)$ \\
$1^2D_{\frac{5}{2}^+}$   ~~&$|70,^{2}10,2,2,\frac{5}{2}^+\rangle$            & $\frac{1}{\sqrt{2}}\left(\phi^{s}\chi^{\rho}\psi_{22L_{z}}^{\rho}+\phi^{s}\chi^{\lambda}\psi_{22L_{z}}^{\lambda}\right)$  \\
$1^4D_{\frac{1}{2}^+}$   ~~&$|56,^{4}10,2,2,\frac{1}{2}^+\rangle$            & $\phi^{s}\chi^{s}\psi_{22L_{z}}^{s}$                                                                                   \\
$1^4D_{\frac{3}{2}^+}$   ~~&$|56,^{4}10,2,2,\frac{3}{2}^+\rangle$            & $\phi^{s}\chi^{s}\psi_{22L_{z}}^{s}$             \\
$1^4D_{\frac{5}{2}^+}$   ~~&$|56,^{4}10,2,2,\frac{5}{2}^+\rangle$            & $\phi^{s}\chi^{s}\psi_{22L_{z}}^{s}$                                                       \\
$1^4D_{\frac{7}{2}^+}$   ~~&$|56,^{4}10,2,2,\frac{7}{2}^+\rangle$            & $\phi^{s}\chi^{s}\psi_{22L_{z}}^{s}$                                                      \\
\hline\hline
\end{tabular}
\end{center}
\end{table}

\begin{table}[htp]
\begin{center}
\caption{\label{Spatialwavefunction} The spatial functions $\psi^{\sigma}_{NLM_L}(\mathbf{p}_\rho, \mathbf{p}_\lambda)$ as the linear combination of $\psi_{n_\rho l_\rho m_\rho}(\mathbf{p}_\rho) \psi_{n_\lambda l_\lambda m_\lambda}(\mathbf{p}_\lambda)$. }
\scalebox{1.0}{
\begin{tabular}{cccccccccccc}\hline\hline
$\psi^{S}_{000}(\mathbf{p}_\rho, \mathbf{p}_\lambda)=\psi_{000}(\mathbf{p}_\rho)\psi_{000}(\mathbf{p}_\lambda)$\\
$\psi^{\rho}_{11M_L}(\mathbf{p}_\rho, \mathbf{p}_\lambda)=\psi_{01M_L}(\mathbf{p}_\rho)\psi_{000}(\mathbf{p}_\lambda)$\\
$\psi^{\lambda}_{11M_L}(\mathbf{p}_\rho, \mathbf{p}_\lambda)=\psi_{000}(\mathbf{p}_\rho)\psi_{01M_L}(\mathbf{p}_\lambda)$\\
$\psi^{S}_{200}(\mathbf{p}_\rho, \mathbf{p}_\lambda)=\frac{1}{\sqrt{2}}[\psi_{100}(\mathbf{p}_\rho)\psi_{000}(\mathbf{p}_\lambda)+\psi_{000}(\mathbf{p}_\rho)\psi_{100}(\mathbf{p}_\lambda)]$\\
$\psi^{\lambda}_{200}(\mathbf{p}_\rho, \mathbf{p}_\lambda)=\frac{1}{\sqrt{2}}[-\psi_{100}(\mathbf{p}_\rho)\psi_{000}(\mathbf{p}_\lambda)+\psi_{000}(\mathbf{p}_\rho)\psi_{100}(\mathbf{p}_\lambda)]$\\
$\psi^{\rho}_{200}(\mathbf{p}_\rho,\mathbf{p}_\lambda)$~~~~~~~~~~~~~~~~~~~~~~~~~~~~~~~~~~~~~~~~~~~~~~~~~~~~~~~~~~~~~~~~~~~~~~~~~~~~~~~~~~~~\\
$=\frac{1}{\sqrt{3}}[\psi_{011}(\mathbf{p}_\rho)\psi_{01-1}(\mathbf{p}_\lambda)-\psi_{010}(\mathbf{p}_\rho)\psi_{010}(\mathbf{p}_\lambda)+\psi_{01-1}(\mathbf{p}_\rho)\psi_{011}(\mathbf{p}_\lambda)]$\\
$\psi^{S}_{22M_L}(\mathbf{p}_\rho, \mathbf{p}_\lambda)=\frac{1}{\sqrt{2}}[\psi_{02M_L}(\mathbf{p}_\rho)\psi_{000}(\mathbf{p}_\lambda)+\psi_{000}(\mathbf{p}_\rho)\psi_{02M_L}(\mathbf{p}_\lambda)]$\\
$\psi^{\lambda}_{22M_L}(\mathbf{p}_\rho, \mathbf{p}_\lambda)=\frac{1}{\sqrt{2}}[\psi_{02M_L}(\mathbf{p}_\rho)\psi_{000}(\mathbf{p}_\lambda)-\psi_{000}(\mathbf{p}_\rho)\psi_{02M_L}(\mathbf{p}_\lambda)]$\\
$\psi^{\rho}_{222}(\mathbf{p}_\rho, \mathbf{p}_\lambda)=\psi_{011}(\mathbf{p}_\rho)\psi_{011}(\mathbf{p}_\lambda)$\\
$\psi^{\rho}_{221}(\mathbf{p}_\rho, \mathbf{p}_\lambda)=\frac{1}{\sqrt{2}}[\psi_{010}(\mathbf{p}_\rho)\psi_{011}(\mathbf{p}_\lambda)+\psi_{011}(\mathbf{p}_\rho)\psi_{010}(\mathbf{p}_\lambda)]$\\
$\psi^{\rho}_{220}(\mathbf{p}_\rho,\mathbf{p}_\lambda)$~~~~~~~~~~~~~~~~~~~~~~~~~~~~~~~~~~~~~~~~~~~~~~~~~~~~~~~~~~~~~~~~~~~~~~~~~~~~~~~~~~~~\\
$=\frac{1}{\sqrt{6}}[\psi_{01-1}(\mathbf{p}_\rho)\psi_{011}(\mathbf{p}_\lambda)+2\psi_{010}(\mathbf{p}_\rho)\psi_{010}(\mathbf{p}_\lambda)+\psi_{011}(\mathbf{p}_\rho)\psi_{01-1}(\mathbf{p}_\lambda)]$\\
$\psi^{\rho}_{22-1}(\mathbf{p}_\rho, \mathbf{p}_\lambda)=\frac{1}{\sqrt{2}}[\psi_{01-1}(\mathbf{p}_\rho)\psi_{010}(\mathbf{p}_\lambda)+\psi_{010}(\mathbf{p}_\rho)\psi_{01-1}(\mathbf{p}_\lambda)]$\\
$\psi^{\rho}_{22-2}(\mathbf{p}_\rho, \mathbf{p}_\lambda)=\psi_{01-1}(\mathbf{p}_\rho)\psi_{01-1}(\mathbf{p}_\lambda)$\\
\hline\hline
\end{tabular}}
\end{center}
\end{table}

%\subsection{Wave functions and parameters}
To calculate $\mathcal{M}^{PC(PV)}$ in Eq.~(\ref{Hwwd}),
the details of the wave functions are required.
The total wave function of a baryon system should include four parts: a color wave function $\zeta$, a flavor wave function $\phi$, a spin wave function $\chi$, and a spatial wave function $\psi$.
The spin wave function of a baryon satisfies SU(2) symmetry, which can be expressed as symmetric ($\chi^S_{S_z}$),  mixed antisymmetric($\chi^\rho_{S_z}$), mixed symmetric($\chi^\lambda_{S_z}$) spin wave functions.
The detailed representations of the spin wave function is shown in Refs.~\cite{Niu:2020gjw,Wang:2017kfr,Xiao:2013xi}.

For the $\Omega$ spectrum, the flavor-spin wave functions are representations of
SU(6), which are denoted by $|N_6, ^{2S+1}N_3 \rangle$, where $N_6$ ($N_3$) represents the SU(6) (SU(3)) representation and $S$ stands for the
total spin of the wave function, and the detailed wave function of $\Omega^{(*)}$ excited states are shown in Table~\ref{wavefunction}.
In momentum space, we present the simple harmonic oscillator wave functions to describe the  $\Omega^{(*)}$ baryon~\cite{Wang:2018hmi}.
The  explicit forms of the spatial wave function $\Psi_{N L M_L}^{\sigma}(\textbf{p}_\rho,\textbf{p}_\lambda)$
in the momentum space, up to the $N$ = 2 shell, can be found in Table~\ref{Spatialwavefunction}.
The superscript $\sigma$ characterizes the same set of quantum numbers $(N,L,M_L)$
arising from different combinations of
the $\rho$ and $\lambda$ oscillation systems~\cite{Karl:1969iz,Zhenping:1991,Xiao:2013xi}.
Specifically, the combinations are defined by
$N=2(n_{\rho}+n_{\lambda})+l_{\rho}+l_{\lambda}$, $L=l_\rho+l_\lambda,...,|l_\rho-l_\lambda|$, and $M_L=m_\rho+m_\lambda$,
where $(n_i,l_i,m_i$) with $i=\rho$ or $\lambda$
are the principal, orbital, and magnetic quantum numbers, respectively.
The internal momenta of the $\rho$- and $\lambda$-oscillators,
$\textbf{p}_\rho$ and $\textbf{p}_\lambda$,
are expressed as
\begin{eqnarray}
\mathbf{p}_\rho&=&\frac{\sqrt{2}}{2}(\textbf{p}_1-\textbf{p}_2)\,,\nonumber\\
\mathbf{p}_\lambda&=&
\frac{\sqrt{6}}{2}\frac{m_5(\textbf{p}_1+\textbf{p}_2)-(m_1+m_2)\textbf{p}_5}{m_1+m_2+m_5}\,,
\end{eqnarray}
respectively.
%The  explicit forms of the spatial wave function $\Psi^{\sigma}_{NLM_L}(\textbf{p}_\rho,\textbf{p}_\lambda)$ up to the $N$ = 2 shell have been given in Table~\ref{Spatialwavefunction}.

In momentum space,  $\psi_{n l m}(\textbf{p})$ is expressed by
\begin{eqnarray}\label{wf}
\psi_{n l m}(\textbf{p})=(i)^l(-1)^n\left[\frac{2n!}{(n+l+1/2)!}\right]^{1/2}\frac{1}{\alpha^{l+3/2}}  \nonumber \\
\mathrm{exp}\left(-\frac{\textbf{p}^2}{2\alpha^2}\right)L_n^{l+1/2}(\textbf{p}^2/\alpha^2)\mathcal{Y}_{lm}(\textbf{p})\,,
\end{eqnarray}
where the $l$th solid harmonic polynomial is defined as
$\mathcal{Y}_{lm}(\textbf{p})=|\textbf{p}|^{l}Y_{lm}(\mathbf{\hat{p}})$,
and the oscillator parameter $\alpha$ can be either $\alpha_\rho$ or $\alpha_\lambda$.
For the $\Omega^{(*)}$ wave function,  we define $\alpha\equiv\alpha_\rho=\alpha_\lambda$.
In this work, we use the $\alpha$ value obtained from single Gaussian fitting in Ref.~\cite{Liu:2019wdr}.

For $\Omega_b$ baryon,
the total wave function can be represented  as
\begin{eqnarray}
\Psi_{\Omega_b}(\mathbf{p}_\rho,\mathbf{p}_\lambda)=\zeta_{\Omega_b} \varphi_{\Omega_b}\chi^\lambda_{s_z}\psi_{000}^S(\textbf{p}_\rho,\textbf{p}_\lambda)\,,
\end{eqnarray}
where $\zeta_{\Omega_b}$, $\chi^\lambda_{S_z}$, and $\psi_{000}^S(\textbf{p}_\rho,\textbf{p}_\lambda)$
represent the color, spin, and spatial wave functions, respectively.
It is worth noting that in momentum space,
we employ the simple harmonic oscillator wave functions to describe  $\psi_{000}^S(\textbf{p}_\rho,\textbf{p}_\lambda)$, which is represented as
\begin{eqnarray}
\psi^{S}_{000}(\mathbf{p}_\rho, \mathbf{p}_\lambda)=\psi_{000}(\mathbf{p}_\rho)\psi_{000}(\mathbf{p}_\lambda)
\end{eqnarray}
where $\textbf{p}_\rho$ and $\textbf{p}_\lambda$ are  the internal momenta of the $\rho$- and $\lambda$-oscillators, respectively.
It is worth mentioning that the parameter $\alpha_\lambda$ of  $\Omega_b$ wave function is given by
$\alpha_{\lambda}=[(3m_b)/(2m_s+m_b)]^{1/4}\alpha_{\rho}$~\cite{Wang:2017kfr,Yao:2018jmc}.

For the wave function of $J/\psi$ meson, which can be represented  as
\begin{eqnarray}
\Psi(\textbf{p}'_3,\textbf{p}_4)=\zeta_{J/\psi}\varphi_{J/\psi}\chi^1_{s_z}\psi(\textbf{p}'_3,\textbf{p}_4)\,,
\end{eqnarray}
where $\zeta_{J/\psi}$ is the color wave function. $\varphi_{J/\psi}$, $\chi^1_{s_z}$, and $\psi(\textbf{p}'_3,\textbf{p}_4)$
represent the flavor, spin, and spatial wave functions, respectively,
which are given as:
\begin{eqnarray}\label{3wf}
&&
\chi^1_{1,0,-1}=
(\uparrow\uparrow,(\uparrow\downarrow+\downarrow\uparrow)/\sqrt 2,\downarrow\downarrow)\,,\nonumber\\
&&
\varphi_{J/\psi}=c\bar{c}\,,\nonumber\\
&&
\psi(\textbf{p}'_3,\textbf{p}_4)=\frac{1}{\pi^{3/4}\beta^{3/2}}\mathrm{exp}
\left[-\frac{(\textbf{p}'_3-\textbf{p}_4)^2}{8\beta^2}\right]\,.
\end{eqnarray}
In Eq.~(\ref{3wf}), $s_z$ denotes the $z$ component of the $J/\psi$ spin and  $\beta$ in $\psi(\textbf{p}'_3,\textbf{p}_4)$
controls the width of Gaussian distribution in momentum space.

To convert ${\cal M}(\Omega_b \to J/\psi \Omega^{(*)})$
from Eqs.~(\ref{Hwwb})~and~(\ref{Hwwc}) into the decay width,
we use the following equation:
\begin{equation}\label{widthfun}
\Gamma=8\pi^2\frac{|\textbf{q}|E_{J/\psi}E_{\Omega^{(*)}}}{M_{\Omega_b}}\frac{1}{2J_{\Omega_b}+1}
\sum_{J_{i}^z,J_{f}^z}|\mathcal{M}_{J_{i}^z,J_{f}^z}|^2\,,
\end{equation}
which is applied in our numerical analysis.
In Eq.~(\ref{widthfun}),  $\mathcal{M}_{J_{i}^z,J_{f}^z}$ is the transition amplitude.

\section{Numerical results}\label{numerical}
For our numerical analysis, we adopt the CKM matrix elements
and the masses of $\Omega_{b} $ and $J/\psi$ from PDG~\cite{pdg},
while the quark masses are taken from~\cite{Wang:2017kfr},
as follows:
\begin{eqnarray}
&&
(V_{cb},V_{cs})=(0.042,0.987)\,,\nonumber\\
&&
(m_{\Omega_b},m_{\Omega},m_{J/\psi})
=(6.046,1.672,3.0969)~{\rm GeV}\,,\nonumber\\
&&
(m_s,m_c,m_b)=(0.45,1.48,5.0)~{\rm GeV}\,.
\end{eqnarray}
The lifetime of the $\Omega_b$ state is also taken from PDG~\cite{pdg}:
$\tau_{\Omega_b}=1.64\times 10^{-12} s$. The Wilson coefficient
$C_2 = -0.365$ is from~\cite{Hsiao:2021mlp,Ali:1998eb,Hsiao:2014mua}.
For the harmonic oscillator parameters,
we use $(\alpha_\lambda,\alpha_{\rho})=(0.56,0.44)$~GeV
for $\Omega_b$~\cite{Zhong:2007gp}, and
$\beta =0.50$~GeV for $J/\psi$, as adopted in~\cite{Barnes:2005pb,Xiao:2018iez}.
For the ground-state and excited $\Omega$ hyperon states,
the masses and the oscillator parameters $\alpha$, calculated in~\cite{Liu:2019wdr},
are summarized in Table~\ref{result}.

%
% =========================================================================
\begin{table}[t]
\begin{center}
\caption{\label{result} Our results for $\Omega_b \to J/\psi \Omega^{(*)}$ using the constituent model,
with the quantum numbers $n^{2S+1}J^{P}$ assigned to $\Omega^{(*)}$.
The parameter $\alpha$ and the mass of $\Omega^{(*)}$ $M_f$ are both given in units of MeV;
$\Gamma$ and ${\cal B}$ are in units of $10^{-17}$~GeV and $10^{-4}$, respectively.}
\begin{tabular}{lccccccccccccccccccccccccccccccccccccccccccccc}\hline\hline
$\Omega^{(*)}$ hyperon
&$\alpha$~\cite{Liu:2019wdr}  &$M_f$~\cite{Liu:2019wdr}        &$\Gamma$   &$\mathcal{B}$
&$\frac{{\cal B}(\Omega_b^-\to J/\psi\Omega^{(*)})}{{\cal B}(\Omega_b^-\to J/\psi\Omega^-)}$
\\\hline
$\Omega(1^4S_{\frac{3}{2}^+})$  &440       &1672 &36                 &8.8                &1.00        \\
$\Omega(1^2P_{\frac{1}{2}^-})$  &428       &1957 &17                 &4.2                &0.48            \\
$\Omega(1^2P_{\frac{3}{2}^-})$  &411       &2012 &42                 &11                 &1.25          \\
$\Omega(2^2S_{\frac{1}{2}^+})$  &387       &2232 &18                 &4.5                &0.51         \\
$\Omega(2^4S_{\frac{3}{2}^+})$  &381       &2159 &79                 &20                 &2.3          \\
$\Omega(1^2D_{\frac{3}{2}^+})$  &394       &2245 &7.5                &1.9                &0.22           \\
$\Omega(1^2D_{\frac{5}{2}^+})$  &380       &2303 &17                 &4.2                &0.48          \\
$\Omega(1^4D_{\frac{1}{2}^+})$  &413       &2141 &9.3                &2.3                &0.26      \\
$\Omega(1^4D_{\frac{3}{2}^+})$  &399       &2188 &21                 &5.4                &0.61       \\
$\Omega(1^4D_{\frac{5}{2}^+})$  &383       &2252 &29                 &7.2                &0.81           \\
$\Omega(1^4D_{\frac{7}{2}^+})$  &367       &2321 &14                 &3.3                &0.38          \\
\hline\hline
\end{tabular}
\end{center}
\end{table}
% ====================================================
%

\begin{figure}[t]
\centering
\includegraphics[width=0.45\textwidth]{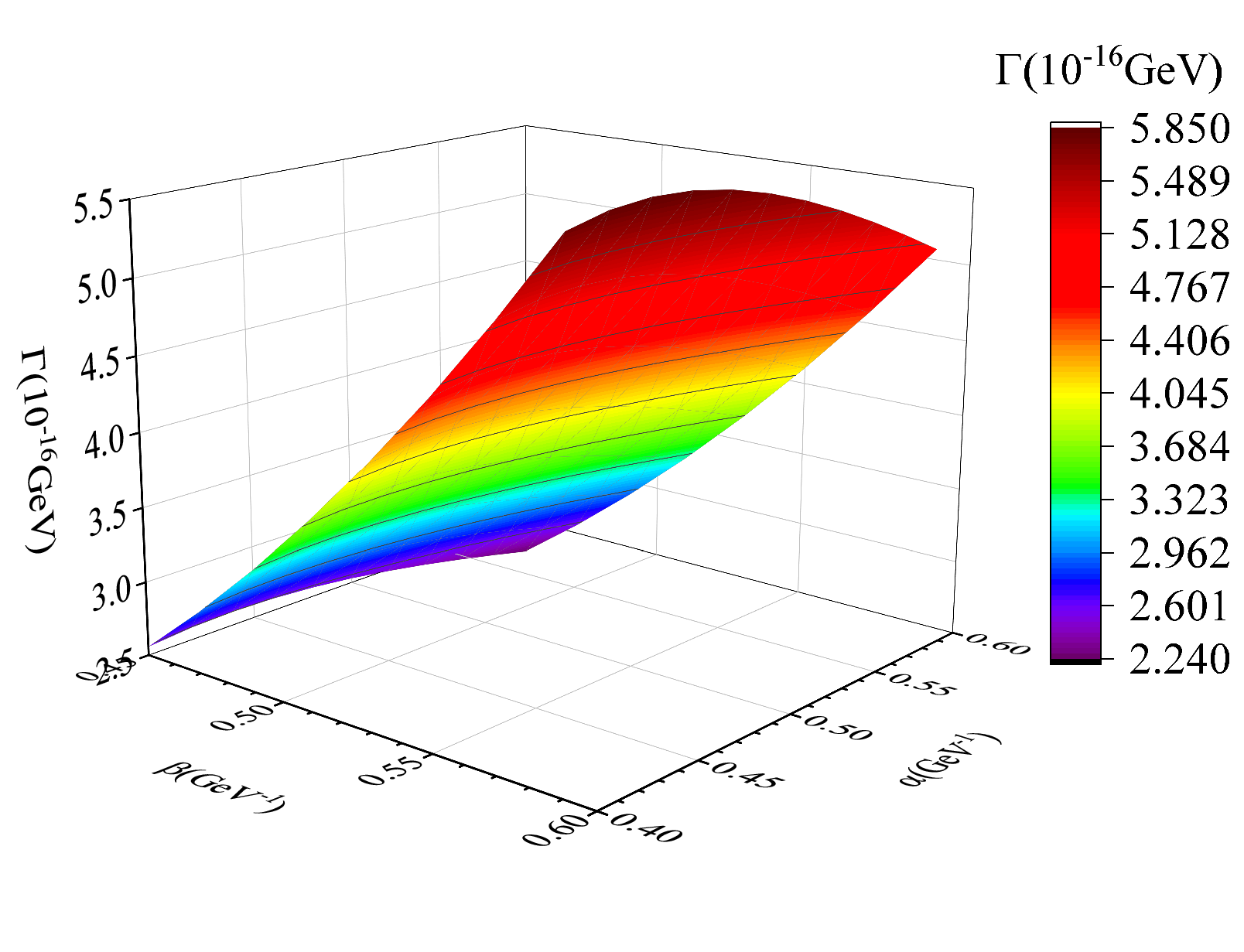} \vspace{-0.5 cm}
\caption{The decay width of $\Omega_b \to J/\psi\Omega $ varies
with the harmonic parameters $\alpha$ of $\Omega$ baryon and
$\beta$ of $J/\psi$ meson.}\label{harmonic parameters}
\end{figure}

Utilizing the inputs above,
we calculate ${\cal B}(\Omega_b \to J/\psi\Omega)$
to compare it with the previous studies. Specifically,
the decay widths and branching fractions of possible
$\Omega_b \to J/\psi\Omega^{(*)}$ are presented in Table~\ref{result}.
Additionally, we plot $\Gamma(\Omega_b \to J/\psi\Omega)$
as functions of the parameters $\alpha$ and $\beta$ in Fig.~\ref{harmonic parameters}
and $\Gamma(\Omega_b \to J/\psi\Omega^{*})$ as a function of $m_{\Omega^*}$ in Fig.~\ref{1Dwave}.
%
% ===============================
\begin{table}[t]
\begin{center}
\caption{\label{s-wave}The branching fraction of $\Omega_b  \rightarrow J/\psi \Omega$
is expressed in units of $10^{-4}$, compared with results from different models.}
\begin{tabular}{lccccccccccccccccccccccccccccccccccccccccccccc}\hline\hline
&Our work&~~\cite{Gutsche:2018utw} &~~\cite{Fayyazuddin:2017sxq}  &~~\cite{Hsiao:2021mlp}  &~~\cite{Cheng:1996cs} &~~\cite{Rui:2023fiz}   &~~\cite{Wang:2024mjw}\\ \hline
 ~~~~&8.8   &8  &0.45    &5.3 &16.7  &6.9$^{+0.5+1.0}_{-0.0-0.3}$  &6.9$^{+2.4}_{-1.7}$ \\
\hline\hline
\end{tabular}
\end{center}
\end{table}
% =============================
%
%
% =============================
% ============================
%
\section{Discussions and conclusion}\label{discussions}
The branching fraction of $\Omega_b \to J/\psi\Omega$,
on the order of $10^{-5}-10^{-3}$, indicates inconclusive calculations
from~\cite{Gutsche:2018utw,Fayyazuddin:2017sxq,Hsiao:2021mlp,Cheng:1996cs,Rui:2023fiz}.
On the other hand, the constituent quark model has been applied to $\Omega_c^0$ decays~\cite{Wang:2022zja},
where ${\cal B}(\Omega_c^0\to\pi^+\Omega(2012)^-)/{\cal B}(\Omega_c^0\to\pi^+\Omega^-)=0.21$
agrees with the experimental value of $0.220\pm 0.059\pm 0.035$~\cite{Belle:2021gtf},
suggesting its applicability to  $\Omega_b$ decays.

In the constituent quark model, we obtain
$\mathcal{B}(\Omega_b \rightarrow \Omega J/\psi)=8.8 \times 10^{-4}$,
which agrees with the calculations using the light-front quark model~\cite{Hsiao:2021mlp,Wang:2024mjw},  covariant confined quark model~\cite{Gutsche:2018utw}
and the perturbative QCD approach~\cite{Rui:2023fiz} to the order of the magnitude,
as displayed in Table~\ref{s-wave}. According to the partial observation~\cite{pdg},
$\emph{f}_{\Omega_b}\mathcal{B}(\Omega_b \rightarrow J/\psi \Omega)
= (1.4_{-0.4}^{+0.5}) \times 10^{-6}$, and with the branching fraction substituted by
our result, we estimate the fragmentation fraction
\emph{f}$_{\Omega_b}\simeq 0.16 \times 10^{-2}$~\cite{Hsiao:2021mlp,Hsiao:2015txa},
which denotes the $b\to\Omega_b$ production rate.

It should be pointed out that our calculation of $\Omega_b \rightarrow \Omega J/\psi$
depends on $\alpha$ and $\beta$. To test the sensitivity, we depict
$\Gamma(\Omega_b \rightarrow J/\psi \Omega)$ as a function of $\alpha$ and $\beta$,
within the parameter space:
$0.4~{\rm GeV}<\alpha<0.5~{\rm GeV}$ and  $0.48~{\rm GeV}<\beta<0.55~{\rm GeV}$.
In Fig.~\ref{harmonic parameters}, it can be seen that
$\Gamma(\Omega_b \rightarrow J/\psi \Omega)$ is more sensitive to $\beta$,
varying from 3.0 to 4.2 times $10^{-16}$~GeV. However,
it remains close to the central value: $3.6 \times 10^{-16}$~GeV~(see Table~\ref{result}).

There are two excited $1P$-wave $\Omega$ states in the $\Omega$ hyperon spectroscopy:
$\Omega(1^2P_{1/2^-})$ and $\Omega(1^2P_{3/2^-})$.
It is interesting to note that the newly observed $\Omega(2012)$ state
is more likely to be assigned to the latter: $\Omega(1^2P_{3/2^-})$,
based on the fact that the measured mass and decay width are consistent with the quark model predictions~\cite{Xiao:2018pwe,Liu:2019wdr,Aliev:2018yjo,Aliev:2018syi,Polyakov:2018mow}.

Based on this assignment, we obtain
${\cal B}(\Omega_b \to J/\psi\Omega(2012))=1.1 \times 10^{-3}$. Specifically,
${\cal B}(\Omega_b \to J/\psi\Omega(2012))/{\cal B}(\Omega_b \to J/\psi\Omega) \simeq 1.3$,
which clearly avoids the uncertain fragmentation fraction $f_{\Omega_b}$,
making it beneficial for experimental examination. In fact,
$\Omega(2012)$ has been observed~\cite{Belle:2018mqs} and reconfirmed
in $\Omega_c^0 \to \pi^+ \Omega(2012)^-,\Omega(2012)^-\to \Xi^0K^-$~\cite{Belle:2021gtf}.
Likewise, this can be reconfirmed from the resonant $\Omega_b$ decays, such as
$\Omega_b \to J/\psi \Omega(2012)$, followed by the subsequent decay
$\Omega(2012) \rightarrow \Xi^0 K^-$.
In particular, $\Omega(2012)\to \Xi K$ has a branching fraction of around~$90\%$~\cite{Zhong:2022cjx}.
The $ \Omega (1^2P_{1/2^-}) $ is a typically missing $\Omega$ hyperon,
with a predicted mass around 1950~MeV~\cite{Engel:2013ig,Capstick:1986ter,Faustov:2015eba,Liu:2019wdr}.
In our evaluation,
${\cal B}(\Omega_b \to J/\psi\Omega(1^2P_{1/2^-}))=4.2 \times 10^{-4}$
indicates that $\Omega(1^2P_{1/2^-})$ has a smaller but compatible production rate
to its $1P$-wave cousin state.
%
% ======================================
\begin{figure}[t]
\centering
\includegraphics[width=0.45\textwidth]{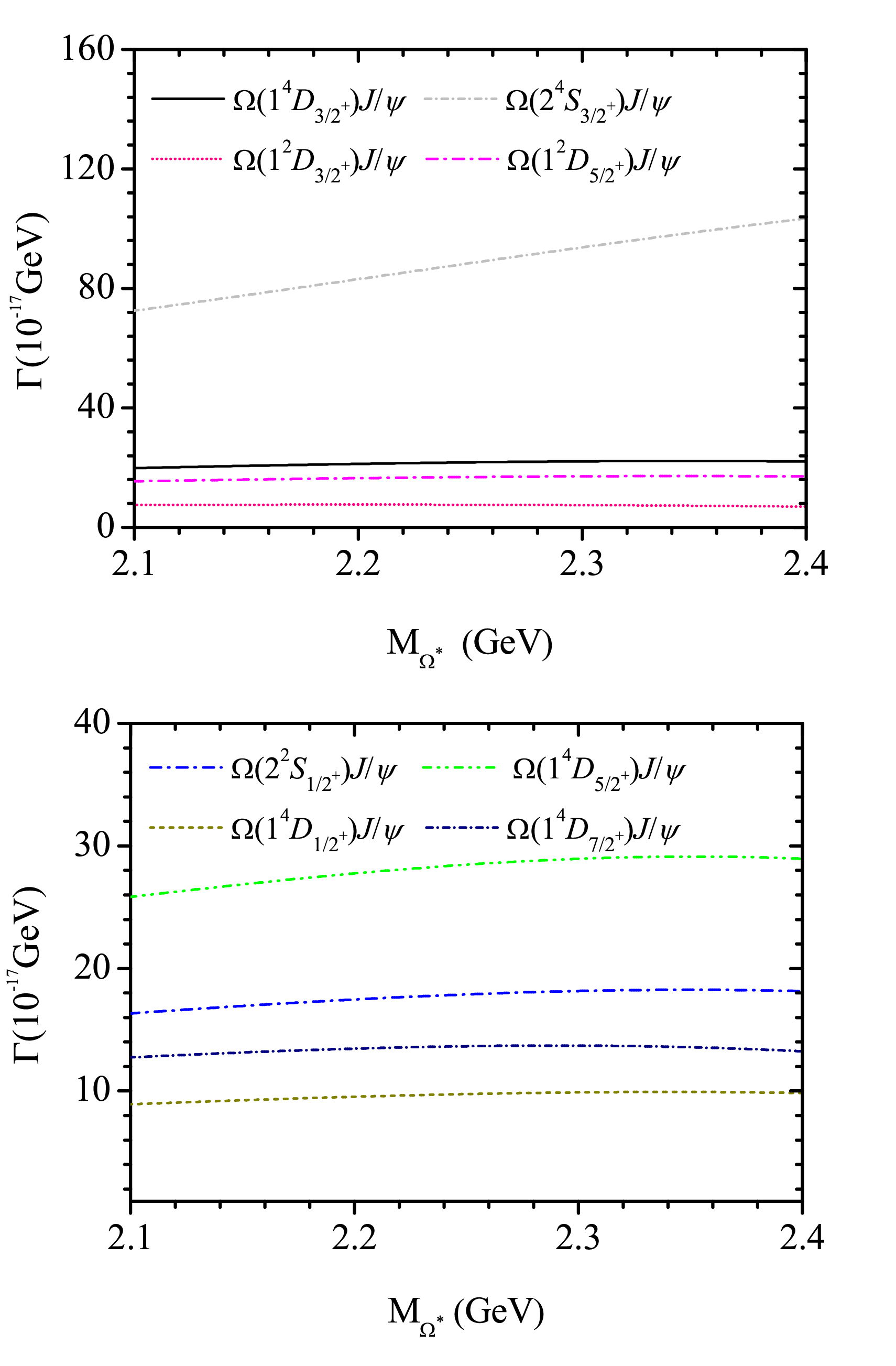} \vspace{-0.5 cm}
\caption{The decay widths of $\Omega_b  \rightarrow J/\psi\Omega^{*}$ as a function of $M_{\Omega^*}$,
with the range from the theoretical estimations.}\label{1Dwave}
\end{figure}
% ======================================
%

In the quark model, the $1D$-wave $\Omega^*$ hyperons are classified into
the spin-doublet $\Omega(1^2D_{3/2^+},1^2D_{5/2^+})$ and
the spin-quartet $\Omega (1^4D_{1/2^+},1^4D_{3/2^+},1^4D_{5/2^+},1^4D_{7/2^+})$,
with masses evaluated to be
around 2150 to 2250~MeV~\cite{Liu:2019wdr,Xiao:2018pwe,Wang:2018hmi,Wang:2022zja}
in the mass spectrum.
However, these states have yet to be firmly established experimentally.

In accordance with the masses from~Ref.~\cite{Liu:2019wdr},
we calculate $\Omega_b \to J/\psi\Omega(1D)$.
Our the results are listed in Table~\ref{result}, where
$\Omega(1^2D_{5/2^+},1^4D_{3/2^+},1^4D_{5/2^+},1^4D_{7/2^+})$ states
have production rates to be 0.5, 0.6, 0.8, 0.4, respectively,
relative to their ground-state counterpart. We highlight that
${\cal B}(\Omega_b \to J/\psi\Omega(1^4D_{5/2^+}))=7.2 \times 10^{-4}$
is sufficiently large for a promising establishment of $\Omega(1^4D_{5/2^+})$
through $\Omega_b \to\Omega(1^4D_{5/2^+}) J/\psi$,
followed by $\Omega(1^4D_{5/2^+}) \to\Xi(1530)^0 K^-$,
where ${\cal B}(\Omega(1^4D_{5/2^+})\to \Xi(1530)K)$
is as large as $80\%$~\cite{Liu:2019wdr,Xiao:2018pwe}.

As for the existing $2S$-wave $\Omega^*$ hyperon states,
$\Omega(2^2S_{1/2^+})$ and $\Omega(2^4S_{3/2^+})$,
we obtain ${\cal B}(\Omega_b \to J/\psi \Omega(2^2S_{1/2^+},2^4S_{3/2^+}))
=(4.5,20)\times 10^{-4}$.
Remarkably, in the current $\Omega$ spectroscopy studies,
$\Omega(2^4S_{3/2^+})$ in $\Omega_b \to J/\psi\Omega^*$ is the only one
that can have a production rate larger than that of its ground state counterpart.

The constituent quark model relies on theoretical inputs from the $\Omega^*$ masses.
Due to the nature of these theoretical estimations, the masses carry some uncertainties.
Therefore, in Figs.~\ref{1Dwave},
we depict the decay widths involving various excited $\Omega$ hyperons as
functions of $M_{\Omega^*}$, with the possible ranges determined by the model calculations.
It is found that the predicted decay widths for both $\Omega^*(1P,1D)$ and $\Omega^*(2S)$
are not sensitive to the uncertainties arising from the mass estimations.

In conclusion, we have calculated the sextet $b$-baryon decays
$\Omega_b \to J/\psi\Omega^{(*)}$ using the constituent quark model.
We found that ${\cal B}(\Omega_b \to J/\psi\Omega)=8.8\times 10^{-4}$,
which is consistent with the previous studies.
With $\Omega(2012)$ identified as $\Omega(1^2P_{3/2^-})$,
we found that ${\cal B}(\Omega_b \to J/\psi\Omega(2012))=1.1\times 10^{-3}$,
which is compatible with ${\cal B}(\Omega_b \to J/\psi\Omega)$. Additionally,
the production rates of the $\Omega(1^2D_{5/2^+},1^4D_{3/2^+},1^4D_{5/2^+},1^4D_{7/2^+})$ states
have been calculated to be 0.5, 0.6, 0.8, 0.4, respectively,
relative to their ground-state counterpart. We also found that
${\cal B}(\Omega_b \to J/\psi\Omega(2^2S_{1/2^+},2^4S_{3/2^+}))=
(4.5,20)\times 10^{-4}$, which is promising for measurement by LHCb.
Since we have demonstrated that our calculations are insensitive to
the parameter inputs and uncertainties arising from the $\Omega^*$ masses,
this provides a suitable test-bed to investigate the $\Omega$ hyperon spectroscopy.

\section*{Acknowledgements}

We are very grateful to the referee for pointing out the incorrect calculation of the color factor, and to Qiang Zhao and Di Wang for their very helpful discussions regarding the color factor.
This work was supported by the National Natural Science Foundation of
China (Grants No.12205026, No.12175065 and No.12235018),
and  Applied Basic Research Program of Shanxi Province,
China under Grant No. 202103021223376.
YKH was supported by NSFC (Grants No.~12175128 and No.~11675030). JW was supported by General Programs of Changzhi College.

%\end{spacing}

\end{document}